\documentclass[sigconf]{acmart}
\AtBeginDocument{%
  }

\copyrightyear{2026}
\acmYear{2026}
\setcopyright{cc}
\setcctype{by}
\acmConference[SIGIR '26]{Proceedings of the 49th International ACM SIGIR Conference on Research and Development in Information Retrieval}{July 20--24, 2026}{Melbourne, VIC, Australia}
\acmBooktitle{Proceedings of the 49th International ACM SIGIR Conference on Research and Development in Information Retrieval (SIGIR '26), July 20--24, 2026, Melbourne, VIC, Australia}
\acmDOI{10.1145/3805712.3808441}
\acmISBN{979-8-4007-2599-9/2026/07}

\begin{document}

\title{Negative Data Mining for Contrastive Learning in Dense Retrieval at IKEA.com}


\author{Eva Agapaki}
\affiliation{
  \institution{IKEA Retail (Ingka Group)}
  \city{Conshohocken, PA}
  \country{USA}
}
\email{eva.agapaki@ingka.com}
\orcid{0000-0002-2962-9203}

\author{Amritpal Singh Gill}
\affiliation{
  \institution{IKEA Retail (Ingka Group)}
  \city{Amsterdam}
  \country{The Netherlands}
}
\email{amritpalsingh.gill@ingka.ikea.com}
\orcid{0009-0004-4399-8680}

\renewcommand{\shortauthors}{Eva Agapaki and Amritpal Singh Gill}


\begin{abstract}
Contrastive learning is a core component of modern retrieval systems, but its effectiveness heavily relies on the quality of negative examples used during training. In this work, we present a systematic approach to improving dense retrieval for IKEA product search through structured negative sampling strategies and scalable LLM-as-a-judge relevance evaluation. 

Building on IKEA Search Engine's late-interaction retrieval architectures, we introduce two key contributions: (1) structured negative sampling strategies that leverage product hierarchical taxonomy and product attributes to generate semantically challenging negatives, and (2) a comprehensive LLM-based evaluation methodology for generating training data. Rather than relying on sparse human annotations or random sampling, our LLM-based evaluation system allocates a score for all candidate products against each query.

Our methodology achieves +2.6\% average category accuracy on offline real user query experiments on the Canada market. However, our A/B test on long-tail queries showed no statistically significant differences in user engagement metrics between the improved and baseline models ($p > 0.05$). We trace this gap to user search behavior: 67\% of popular searches exhibit zero-click rates above 50\%, indicating that a substantial proportion of search sessions result in no product engagement regardless of result ranking. These findings underscore the importance of hard negative mining but also the need for grounding training data and offline evals in real user search behavior---including query intent distribution and zero-click patterns---to bridge the gap between offline retrieval quality and online user engagement.

\end{abstract}



\maketitle

\section{Introduction}

Dense retrieval models have emerged as a promising approach for e-commerce product search, offering the ability to capture semantic relationships between user queries and product descriptions beyond traditional keyword-based matching \cite{colbertv2, karpukhin2020dense}. These models are typically trained using contrastive learning with hard negative mining strategies, where careful selection of challenging, semantically similar examples has shown to yield substantial improvements on offline evaluation benchmarks \cite{kalantidis2020hard, moreira2024nvretriever}. Yet a fundamental question remains: \textit{do these offline gains translate to improved user experiences and business outcomes?}

The relationship between offline retrieval metrics and online performance has been the subject of considerable debate. \cite{wang2023amazon} conducted a comprehensive analysis of 36 offline metrics for product ranking at Amazon, finding that offline metrics align well with online business metrics—agreeing on which of two ranking models performs better up to 97\% of the time—with NDCG exhibiting discriminative power exceeding 99\%. However, this strong directional agreement does not guarantee that offline improvements translate proportionally to online gains. Research at AliExpress \cite{huzhang2021aliexpress} found that there might be significant inconsistency between offline metrics and online performance, attributing this to two factors: (1) offline evaluation ignores item context effects such as position bias and mutual influences among ranked products, and (2) offline datasets provide insufficient signal for learning contextual relationships that matter in production. Their proposed context-aware approach achieved >2\% improvement in conversion rate over models optimized with traditional offline metrics. These findings suggest that while offline metrics capture relevance quality offline, they do not fully reflect user satisfaction in more complex search scenarios.   

This gap between offline and online performance is particularly relevant to dense retrieval for e-commerce product search. Our prior deployment of a late-interaction semantic search model \cite{amrit2025} illustrated this gap: while the model showed significant improvements on long-tail queries, it lacks generalization on attribute-based queries (e.g. "\textit{queen bed mattress}" vs "\textit{king bed mattress}") that reflect patterns of user query behavior.

In this paper, we investigate the end-to-end impact of hard negative mining strategies for dense product retrieval, from offline training to online deployment. We address three research questions:

  \begin{itemize}         
      \item \textbf{RQ1:} How do different hard negative mining strategies affect dense retrieval performance for product search?                            
      \item \textbf{RQ2:} What factors contribute to the generalization gap between synthetic and real user queries?                                                     
      \item \textbf{RQ3:} How do offline retrieval metrics correlate with online user engagement and purchase intent?                                                      
  \end{itemize}                                                                        
We present a comprehensive study using our e-commerce product catalog in the Canada market, comparing multiple training configurations including our baseline random negatives, rule-based hard negatives, and LLM-generated query-centric negatives. Our evaluation framework is based on: (1) a synthetic user query offline benchmark, (2) real user query offline benchmark, and (3) online A/B testing.      

Our findings reveal a consistent pattern: \emph{hard negative mining strategies yield gains in offline experiments} (both on synthetic and real evaluation queries), and do not translate to statistically significant differences in online A/B test metrics. This offline--online gap--consistent with prior observations in e-commerce search \cite{huzhang2021aliexpress}---suggests that \emph{substantial offline improvements do not guarantee online gains}, and that the choice of evaluation queries (synthetic vs.\ real) significantly affects conclusions about model quality. We discuss potential explanations for this gap and provide recommendations for more robust evaluation practices in e-commerce retrieval.

\section{Dataset and Problem Analysis}

\subsection{IKEA Product Data}
We use the IKEA Canada product data that has 24,350 products across 373 leaf product categories (product categories that do not have children categories under them). Each product contains structured metadata including product name (e.g., ``KALLAX''), product category hierarchy, price and attributes such as color, material, and dimensions. Additionally, each product has an LLM-generated product description summarizing its key features, averaging 150 tokens in length according to our prior work \cite{amrit2025}.

\subsection{User Query Analysis}

We analyzed 204,528 unique search queries from 28 days of production search logs in the Canadian market. We classified queries into four types based on detected attributes:

\begin{itemize}                                                                                        \item \textbf{Product name queries} (18.2\%): Queries containing IKEA product series names (e.g., ``kallax'', ``billy bookcase'', ``malm dresser'')               
    \item \textbf{Multi-attribute queries} (12.4\%): Queries with two or more attributes such as color, material, or size (e.g., ``white wooden desk'', ``grey velvet 
  sofa'')                                                               
    \item \textbf{Single-attribute queries} (24.7\%): Queries with exactly one attribute (e.g., ``black shelf'', ``large mirror'')                           
    \item \textbf{Category-only queries} (44.7\%): Generic product type queries without specific attributes (e.g., ``desk'', ``bookshelf'', ``curtains'')             
  \end{itemize}

This distribution reveals that nearly half of user queries are category-only searches, while attribute-based (single- and multi-attribute) queries represent a significant portion (37.1\%), where semantic understanding of query attributes is important. 
\section{Training Data Generation Pipeline}

\begin{figure}[htbp] 
      \centering              
      \includegraphics[width=9cm]{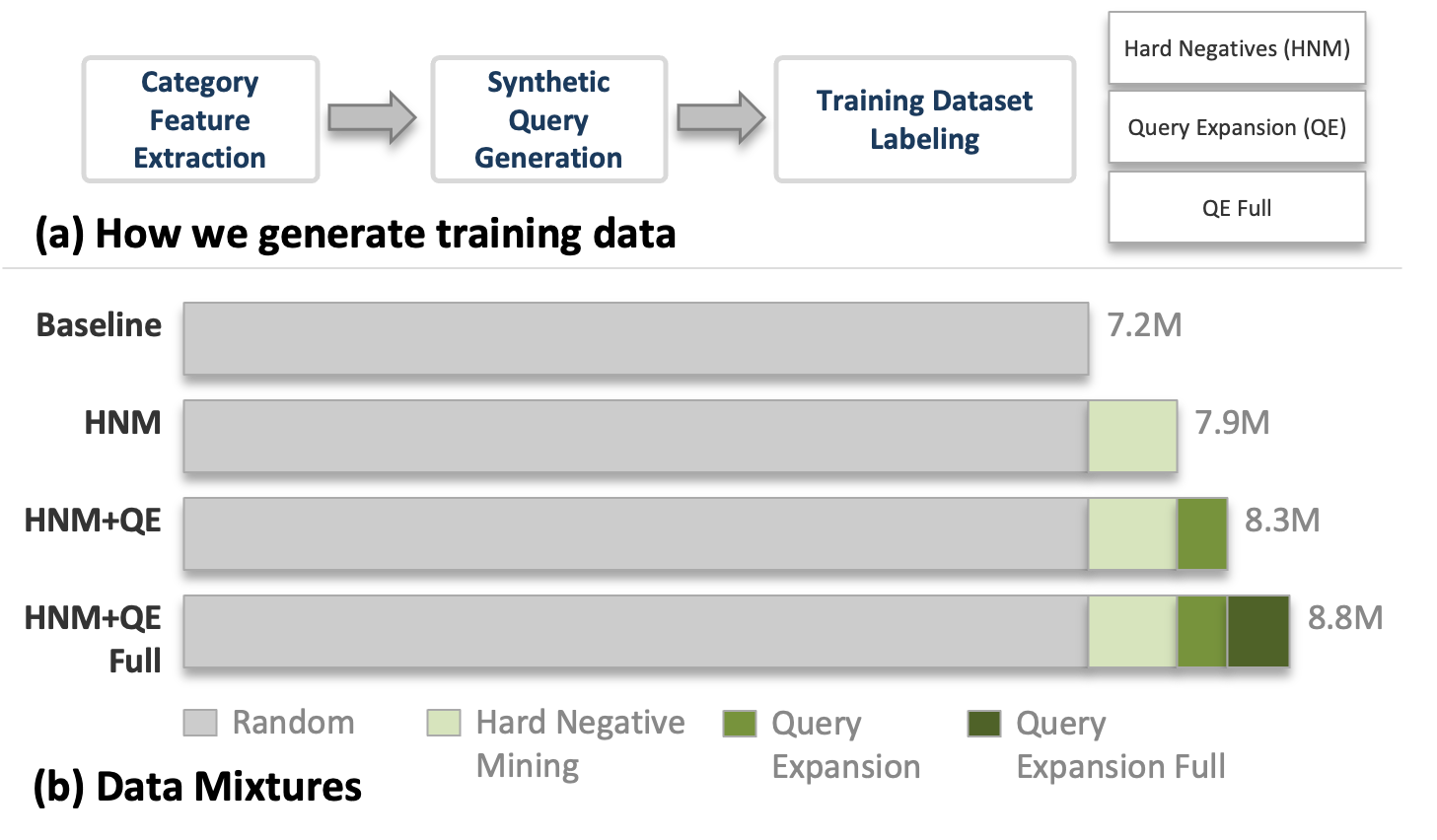}
      \caption{(a) Training data generation pipeline consisting of three stages and (b) data mixtures visualization.}            
      \label{fig:pipeline}              
  \end{figure}

We develop an automated pipeline that generates training triplets by combining LLM-based relevance annotation with structured hard negative mining strategies (Figure~\ref{fig:pipeline}). This approach enables scalable generation of hard negative examples while maintaining label quality through structured evaluation. 

\subsection{Category Feature Extraction}                                        
The first stage uses an LLM to extract category-specific and semantically meaningful attributes from the product catalog rather than relying on predefined attribute vocabularies. For each of the 373 leaf categories, we provide the LLM with a representative sample of products including names, descriptions, common search terms and category metadata. We prompt the LLM to identify visual (e.g., colors), material, size or physical attributes, life at home activities (e.g., cook, sleep) or user needs. This ensures that extracted attributes reflect genuine category semantics---for instance, extracting material attributes like ``solid wood'' and ``veneer'' for furniture but not for textiles---eliminating illogical pairings that arise from keyword-based matching approaches.                              

\subsection{Synthetic Query Generation}                                         

The second stage generates diverse search queries for each category using the extracted attributes as input. We prompt an LLM to generate 50 search queries for each product category across complexity tiers that reflect the natural variation in how users search for products: (1) \textbf{Simple} (30\%): 1--2 words (e.g., ``grey sofa''), (2) \textbf{Medium} (35\%): 3--4 words combining attributes (e.g., ``grey velvet corner sofa'') and (3) \textbf{Complex} (35\%): 5+ words (e.g., ``small grey corner sofa bed with storage''). Beyond attribute-based queries, we generate intent-based variations: \textit{activity-based} (``sofa for watching TV''), \textit{situational} (``sofa for small apartment''), and \textit{product-name} queries (\textit{``KIVIK 3-seat sofa''}). This yields 6,300 unique queries across all categories.
  
\subsection{Training Dataset Labeling}                                                             

Since large-scale manual relevance annotation is expensive, we use LLMs to score query-product relevance on a 1-5 scale \cite{Sachdev2025Automated}. Training triplets are constructed by pairing each query with positive products (LLM-generated score $\geq 4$) and negative products (LLM-generated score $\leq 2$). Products with scores of 3 are excluded to maintain clear separation between positive and negative examples. This labeling process produces 448,073 triplets, which are subsequently classified into subdatasets based on the negative type:                                                                   
  \begin{itemize}                                                                                       
      \item \textbf{Attribute-only} (244K): Same category, different attribute (e.g., query ``white sofa,'' negative is black sofa)                                       
      \item \textbf{Cross-category} (105K): Matching attribute, wrong category (e.g., query ``white sofa,'' negative is white table)                                   
      \item \textbf{Multi-attribute} (93K): Queries with multiple attributes where each negative violates exactly one attribute. Given the query \textit{``white metal outdoor lamp''}, example negatives are white metal \textit{indoor} lamps (wrong setting) or white \textit{wooden} outdoor lamps (wrong material).                                                       
  \end{itemize}                                                                                   
  
This classification enables targeted analysis of model performance on different hard negative types and supports ablation studies on training data composition.

\subsection{Data Mixtures}

We experiment with different compositions of training data to understand the impact of hard negatives on model performance. Our training data mixtures include:     

\begin{itemize}                                                            
    \item \textbf{Baseline}: Random negative sampling of the product catalog, resulting in 7.2M triplets where negatives are uniformly sampled regardless of semantic similarity to the query or positive product.                                                                           \item \textbf{HNM (Hard Negative Mining)}: Combines the baseline with additional 448K LLM-labeled hard negative triplets distributed across negative types: attribute-only (244K), category mismatch (145K), and multi-attribute (311K). Total: 7.9M triplets.     
    \item \textbf{HNM + QE (Query Expansion)}: Extends HNM with additional LLM-generated queries targeting underrepresented query patterns (e.g., query ''white bookshelf``, negative: ``white KALLAX Insert with door'', query ''KALLAX``, negative: ``BILLY Bookcase''), adding 450K triplets focused on activity-based queries, living situation constraints, and IKEA product name variations. Total: 8.3M triplets. \item \textbf{HNM + QE Full}: Further expands with an exhaustive query expansion across all categories, adding 897K triplets. Total: 8.8M triplets.                 
  \end{itemize}                                                                                    
  
All models share the same base architecture and training hyperparameters according to \cite{amrit2025}, differing only in training data composition. This controlled setup enables direct comparison of how training data quality (hard negatives) versus quantity (expanded queries) affects retrieval performance.

\section{Experimental Setup}
\subsection{Models and Training Data}

We fine-tune the same late interaction-based semantic search model using our baseline model's setup from prior work~\cite{amrit2025}. All models are trained with identical hyperparameters: learning rate of $3 \times 10^{-6}$, batch size of 32, and maximum document length of 512 tokens. Training proceeds for a maximum of 500,000 steps with early stopping based on validation loss.                                                                                                           
The late interaction architecture computes query-document similarity through token-level maximum similarity aggregation:                                              
  \begin{equation}                                                                  
      S(q, d) = \sum_{i \in |E_q|} \max_{j \in |E_d|} E_{q_i} \cdot E_{d_j}^T 
   \end{equation}                                                                                                                                                        
where $E_q$ and $E_d$ are the contextualized token embeddings for query and document respectively.                                                 We train four model variants corresponding to the data mixtures described in Section 3.4: Baseline, HNM, HNM+QE, and HNM+QE (Full). Training was conducted on NVIDIA A100 GPUs, with each configuration requiring approximately 18 hours to complete. To ensure statistical robustness, each configuration was trained five times with different random seeds controlling weight initialization and data shuffling, and we report mean $\pm$ standard deviation across seeds in Table~\ref{tab:synth_real_comparison}.

\subsection{Evaluation Framework}

We evaluate models using two complementary offline benchmarks and one online evaluation. For both offline benchmarks, ground truth relevance is determined by LLM scoring on a 1--5 scale, with products scoring $\geq 4$ labeled as positive. To validate the reliability of LLM-generated labels, we conducted a human evaluation on the real query benchmark and found 93\% agreement between human annotation and LLM relevance judgments, indicating that offline evaluation reflects human notions of relevance.

\paragraph{Synthetic Query Benchmark.} We construct 70 evaluation queries using the same LLM-based generation process as training but with held-out product categories spanning simple (30\%), medium (35\%), and complex (35\%) patterns. This benchmark measures in-distribution performance on query patterns similar to training data.                                                                          

\paragraph{Real Query Benchmark.} We sample 100 queries from production search logs, stratified by query type: 35 single-attribute, 35 multi-attribute, and 30 product-name queries. This benchmark measures out-of-distribution generalization to real user search behavior.                          

\paragraph{Online A/B Test.} We deploy the selected model to production and measure user engagement metrics against the baseline model over a 2-week period with 50/50 traffic split. The online test was focused on long-tail queries ($\geq 3$ words).

\section{Results}
\subsection{Evaluation Metrics}                                                                                                                                         
We evaluate model performance using standard retrieval metrics alongside a category accuracy metric designed to measure the relevant product category even if the product is not an exact match.                                                                                                                                                        
\textbf{Category Accuracy@K (Cat@K)} measures the proportion of top-K retrieved products belonging to correct product categories:                                       
  \begin{equation}                                                                                                                                                        
  Cat@K = \frac{|\{p \in \text{Retrieved}@K : \text{cat}(p) \in C_q\}|}{K}                                                                                                
  \end{equation}                                                                                                                                                          
where $C_q$ is the set of leaf categories (according to IKEA's hierarchical product category data) from relevant ground truth products for query $q$. This metric captures partial retrieval success, returning products of the correct type even when exact matches are missed.                                                                               

\subsection{Offline Evaluation Results}
  
Table~\ref{tab:synth_real_comparison} presents the evaluation results on both the synthetic and the real query datasets. 

\begin{table}[H]
    \centering
    \caption{Model performance on synthetic and real query evaluation sets}
    \label{tab:synth_real_comparison}
    \resizebox{\columnwidth}{!}{%
    \begin{tabular}{l|ccc|ccc}
    \toprule
    & \multicolumn{3}{c|}{\textbf{Synthetic (n=70)}} & \multicolumn{3}{c}{\textbf{Real (n=100)}} \\
    \textbf{Model} & R@10 & Cat@10 & Cat@50 & R@10 & Cat@10 & Cat@50 \\
    \midrule
    Baseline & 44.9{\tiny$\pm$0.1} & 76.5{\tiny$\pm$0.6} & 47.5{\tiny$\pm$0.1} & 41.2{\tiny$\pm$0.9} & 73.4{\tiny$\pm$0.4} & 62.1{\tiny$\pm$0.4} \\
    +HNM & 49.1{\tiny$\pm$0.4} & 80.8{\tiny$\pm$0.3} & 48.7{\tiny$\pm$0.2} & 41.2{\tiny$\pm$0.4} & 76.0{\tiny$\pm$0.7} & 64.6{\tiny$\pm$0.3} \\
    +HNM+QE & 48.2{\tiny$\pm$0.5} & 79.8{\tiny$\pm$1.0} & 48.1{\tiny$\pm$0.5} & 38.3{\tiny$\pm$0.9} & 76.0{\tiny$\pm$1.0} & 64.2{\tiny$\pm$0.3} \\
    +HNM+QE Full & 47.6{\tiny$\pm$0.4} & 79.8{\tiny$\pm$0.4} & 48.9{\tiny$\pm$0.5} & 39.8{\tiny$\pm$0.9} & 74.1{\tiny$\pm$0.7} & 63.6{\tiny$\pm$0.3} \\
    \bottomrule
    \end{tabular}%
    }
  \end{table}

The results reveal three key findings. First, hard negative mining improves category accuracy across both benchmarks: the HNM model achieves the highest category accuracy (Cat@10 = 80.8\% on synthetic queries, +4.3\% over baseline) with corresponding recall gains (R@10 = 49.1\%, +4.2\%). However, these improvements diminish on real queries: while HNM improves R@10 by +4.2\% over baseline on synthetic queries, R@10 remains unchanged. This suggests that HNM optimizes effectively for synthetic query patterns but faces challenges generalizing to real queries.

Second, a key factor driving this generalization gap is the difference in query specificity: synthetic queries target an average of 1.7 product categories, while real user queries match products across 12.5 categories---a 7$\times$ difference. Real users often search with broad intent (e.g., ``sofa'') where multiple subcategories are acceptable, whereas synthetic queries are narrowly specific (e.g., ``white leather 3-seat corner sofa''). Notably, Cat@50 improves substantially on real queries compared to synthetic ones (e.g., HNM: 48.7\% $\rightarrow$ 64.6\%), indicating that models retrieve relevant categories but rank them lower when query intent is broad.

Third, models trained with additional LLM-generated data (QE variants) do not outperform HNM on either dataset despite containing up to 900K more training triplets (8.8M vs 7.9M), suggesting that the query expansions inherit the narrow intent distribution of synthetic data and do not improve real-query generalization.
  
\subsection{A/B Test Results}
  
Based on these offline results, we selected the \textbf{HNM} model for the production A/B test. The selection criteria prioritized category accuracy over recall, motivated by the hypothesis that users prefer search results containing products from the correct category, even if not exact matches, over results that include irrelevant product types.  

The results showed no statistically significant differences between the HNM and baseline models across all user engagement metrics, including click-through rate, add-to-cart rate, and search interaction rate ($p > 0.05$). This finding directly addresses RQ3: the +2.6\% average category accuracy improvement observed offline did not translate to measurable changes in user behavior.                                                                                           
\section{Discussion and Conclusion}

This paper investigated the end-to-end impact of hard negative mining strategies for dense product retrieval, from training data generation through offline evaluation to online deployment. For \textbf{RQ1}, hard negative mining yields meaningful offline improvements in category accuracy, but additional training data through query expansion provides diminishing returns --the largest model (HNM+QE Full, 8.8M triplets) underperforms the simpler HNM model (7.9M triplets), indicating that data quality outweighs data quantity. For \textbf{RQ2}, we identify a substantial generalization gap driven by differences in intent breadth: real user queries span 7$\times$ more product categories than synthetic queries. This is consistent with \cite{rahmani2025bias}, who demonstrate that LLM-generated evaluation data exhibits systematic biases including leniency in relevance scoring. Models optimized on narrowly-specific synthetic hard negatives learn fine-grained distinctions that penalize products real users would find acceptable, and LLM-generated query expansions inherit this narrow intent distribution, explaining why training with more data degrades real-query performance.

For \textbf{RQ3}, our A/B test showed no statistically significant differences across engagement metrics, aligning with evidence that offline retrieval gains do not reliably predict online improvements \cite{wang2023amazon, huzhang2021aliexpress}. Our analysis of user search behavior explains this result: 44.7\% of queries are broad category searches where the baseline already performs well, and 67\% of promoted popular searches exhibit zero-click rates above 50\%---indicating that users frequently satisfy their needs from the search results page without clicking \cite{ye2022zeroclicks}. When the majority of sessions end without clicks regardless of ranking quality, retrieval improvements have limited room to influence engagement metrics. Additionally, the restriction to long-tail queries ($\geq 3$ words) limited statistical power; \cite{somanchi2023heterogeneity} show that heterogeneous treatment effects appear in 10--20\% of experiments and average treatment effects mask sub-group differences, while \cite{bi2022interleaving} report that standard A/B tests lack sensitivity for detecting ranking improvements compared to interleaving experiments. These deployment challenges are not unique to our setting: \cite{magnani2022walmart} and \cite{lin2024walmart} report that embedding-based retrieval at Walmart required additional relevance filtering to address false positives invisible in offline evaluation, and \cite{li2021taobao} identified training-inference inconsistencies requiring separate relevance control at serving time.

Our work contributes a controlled comparison of hard negative mining strategies across three evaluation stages---synthetic queries, real queries, and online deployment---revealing that each stage yields different conclusions about model quality. We quantify the generalization gap and document an inconclusive A/B test result with accompanying user behavior analysis, contributing to the underrepresented literature on deployment outcomes essential for calibrating expectations in applied information retrieval. 

Future work should explore training strategies that better reflect the broad intent distribution of real search traffic, including data augmentation through query rewording and cross-market training data to improve coverage and generalization. On the evaluation side, click-based metrics may be insufficient for measuring retrieval quality when the majority of search sessions end without clicks. Alternative engagement signals---such as scroll depth, wishlist additions from search results, filter and sort interactions after search, and query reformulation rates---could provide finer-grained measures of user satisfaction. Combining these signals with offline retrieval metrics in our evaluation framework could help bridge the gap between offline quality and online user impact.

\section*{Bio}
Eva Agapaki is a Sr Data Scientist at IKEA, where she develops large-scale neural retrieval and recommendation systems for e-commerce. Her research focuses on dense retrieval, contrastive learning, and LLM-based evaluation for search relevance. Her work bridges research experimentation and production, advancing practical evaluation methodologies for semantic search at scale. She holds a Ph.D. in Computer Science from the University of Cambridge and MIT.

\section*{Acknowledgments}                                    We thank the Search Engineering and Data Science teams at IKEA Retail (Ingka Group) for their support throughout this work. We are particularly grateful to Fernando Dorado Rueda for MLOps support and production model deployment, and to Anders Walle for defining the product vision and prioritizing search relevance improvements that motivated this research. 

The authors acknowledge the peoples of the Woi Wurrung and Boon Wurrung language groups of the eastern Kulin Nation on whose unceded lands ACM SIGIR 2026 was hosted. We pay our respects to their Elders past and present, and extend that respect to all Aboriginal and Torres Strait Islander peoples today and their continuing connection to land, sea, sky, and community.

\bibliographystyle{ACM-Reference-Format}
\balance
\bibliography{sample-base}

\end{document}